\documentclass[pre,twocolumn,aps,showpacs]{revtex4}

\begin{document}
\centerline{\bf PHYSICAL REVIEW E, v. 59, 3008 (1999)}
\bigskip
\bigskip
\title{Anomalous scalings and dynamics of magnetic helicity}
\author{Igor Rogachevskii}
\email{gary@menix.bgu.ac.il} \homepage{http://www.bgu.ac.il/~gary}
\author{Nathan Kleeorin}
\email{nat@menix.bgu.ac.il}
\affiliation{Department of Mechanical
Engineering, The Ben-Gurion
University of the Negev, \\
POB 653, Beer-Sheva 84105, Israel}
\date{Received 8 June 1998}
\begin{abstract}
It is demonstrated that the two-point correlation function of the
magnetic helicity in the case of zero mean magnetic field has
anomalous scalings for both, compressible and incompressible
turbulent helical fluid flow. The magnetic helicity in the limit
of very high electrical conductivity is conserved. This implies
that the two-point correlation function of the conserved property
does not necessarily have normal scaling. The reason for the
anomalous scalings of the magnetic helicity correlation function
is that the magnetic field in the equation for the two-point
correlation function of the magnetic helicity plays a role of a
pumping with anomalous scalings. It is shown also that when
magnetic fluctuations with zero mean magnetic field are generated
the magnetic helicity is very small even if the hydrodynamic
helicity is large. Astrophysical applications of the obtained
results are discussed.
\end{abstract}
\maketitle

\section{Introduction}

Problems of intermittency and anomalous scalings for vector
(magnetic) and scalar fields passively advected by a turbulent
fluid flow are a subject of active research in the last years
(see, e.g., \cite{ZMR88,K94,CF95,V96,BY96,EKRR96,RK97,EKRR97}).
The anomalous scaling means the deviation of the scaling exponents
of the correlation function of a vector (scalar) field from their
values obtained by the dimensional analysis. An interesting
question is the role of the conservation laws in the problem of
intermittency and anomalous scalings. For the passive scalar
advected by incompressible and homogeneous turbulent fluid flow
the quantity $ n^{2} $ (or $ T^{2}) $ is  conserved (for
infinitely small diffusivity or thermal conductivity), where $ n $
is the number density of particles, and $ T $ is the fluid
temperature. Corresponding two-point correlation function $
\langle n(t,{\bf x}) n(t,{\bf y}) \rangle $ has normal scaling
(see,e.g., \cite{K94,CF95}). For the passive vector (magnetic
field) the quantity $ h^{2} $ is not  conserved and the second
moment $ \langle {\bf h}(t,{\bf x}) \cdot {\bf h}(t,{\bf y})
\rangle $ has anomalous scalings \cite{V96,RK97}, where $ {\bf h}
$ is the magnetic field. In this case the total (magnetic plus
hydrodynamic) energy is  conserved. On the other hand, the
magnetic helicity $ \langle {\bf a}(t,{\bf x}) \cdot {\bf
h}(t,{\bf x}) \rangle $ in the limit of very high electrical
conductivity is conserved. Here $ {\bf a} $ is a vector potential
of magnetic field, i.e., $ {\bf h} = {\bf  \nabla} {\bf \times}
{\bf a} .$ What is a scaling for the two-point correlation
function $ \langle {\bf a}(t,{\bf x}) \cdot {\bf h}(t,{\bf y})
\rangle $ of the magnetic helicity?

In this Letter we show that the two-point correlation function of
magnetic helicity has anomalous scalings for both, compressible and
incompressible turbulent helical fluid flow. For the helical fluid
velocity field $ \alpha(r) \not= 0 $ [see Eq. (\ref{B1}) below].
The reason for the anomalous scalings is that the magnetic
field in the equation for the two-point correlation function of
magnetic helicity plays a role of a pumping with anomalous scalings.
This implies that the two-point correlation function of the conserved
property does not necessarily have normal scaling. This
demonstrates a difference between passive scalar and vector fields.
We also study here an excitation of the magnetic
helicity by a helical turbulent fluid flow in the case of generation of
magnetic fluctuations with zero mean magnetic field.
Note that for a nonhelical velocity field $ [\alpha(r) = 0 ,$ see
Eq. (\ref{B1}) below] the scaling exponent of the magnetic helicity
correlation function is normal, i.e., it can be obtained from the
dimensional arguments \cite{BY96}.

\section{Governing equations}

We study the evolution of magnetic fluctuations with zero mean
magnetic field in a low-Mach-number compressible turbulent fluid
flow. A mechanism of the generation of magnetic fluctuations with
a zero mean magnetic field was proposed in \cite{VZ72} and
comprises stretching, twisting and folding of the original loop of
a magnetic field. These non-trivial motions are three dimensional
and result in an amplification of the magnetic field. The magnetic
field is determined by the induction equation $ \partial {\bf h} /
\partial t  + ({\bf u} \cdot {\bf  \nabla}) {\bf h} = ({\bf h}
\cdot {\bf  \nabla}) {\bf u} - {\bf h} ({\bf  \nabla} \cdot {\bf
u}) + \eta \Delta {\bf h} ,$ where $ {\bf u} $ is the fluid
velocity, $ \eta $ is the magnetic diffusion. We derive equations
for the second-order correlation functions of the magnetic field
and the magnetic helicity and we use a method of path integrals
and modified Feynman-Kac formula (see, e.g.,
\cite{ZMR88,EKRR96,RK97,EKRR97,EKR96}). The equation for the
second-order correlation function $ h_{ij} = \langle h_{i}(t,{\bf
x}) h_{j}(t,{\bf y}) \rangle $ of the magnetic field is
\begin{eqnarray}
\partial h_{ij} / \partial t = [ \hat L_{ik}({\bf x})
\delta_{js} + \hat L_{js}({\bf y}) \delta_{ik} + \hat N_{ijks} ] h_{ks}
+ I_{ij} \;
\label{D30}
\end{eqnarray}
(for details see \cite{RK97}), where
\begin{eqnarray*}
\hat L_{ij} = \varepsilon_{iks} {\partial \over \partial x_k}
\biggr[ \varepsilon_{smj} U_{m} + \alpha_{sj}
- \hat \eta_{sm} \varepsilon_{mpj}
{\partial \over \partial x_p} \biggr] \;,
\\
{1 \over2} \hat N_{ijks} = \delta_{ik} \delta_{js} f_{mn}
{\partial^{2} \over \partial x_m \partial y_{n} }
+ {\partial^{2} f_{ij} \over \partial x_k \partial y_{s} }
- \delta_{ik} {\partial f_{mj} \over \partial y_{s}}
{\partial \over \partial x_m}
\\
-\delta_{js} {\partial f_{in} \over \partial x_{k}}
{\partial \over \partial y_n}
+ \delta_{ik} \delta_{js} {\partial f_{mp} \over \partial y_{p}}
{\partial \over \partial x_m}
+ \delta_{ik} \delta_{js} {\partial f_{pn} \over \partial x_{p}}
{\partial \over \partial y_n}
\\
- \delta_{ik} {\partial^{2} f_{pj} \over \partial x_p \partial y_{s}}
- \delta_{js} {\partial^{2} f_{ip} \over \partial x_k
\partial y_{p} }
+ \delta_{ik} \delta_{js} {\partial^{2} f_{pl} \over \partial x_p
\partial y_{l} } \;,
\end{eqnarray*}
and $ f_{mn} = \langle \tau u_{m}({\bf x}) u_{n}({\bf y}) \rangle ,$
and $ \delta_{mn} $ is the
Kronecker tensor and $ \varepsilon_{ikm} $ is the Levi-Civita tensor,
$ \hat \eta_{ij} = (\eta_{pp} \delta_{ij} - \eta_{ij}) / 2 ,$
and $ {\bf U} = {\bf V} - {\bf \nabla}_{p} \langle \tau u_p
{\bf u} \rangle / 2 ,$ and
$ \alpha_{mn} = -( \varepsilon_{mji} \langle \tau u_i {\bf  \nabla}_{n}
u_j \rangle + \varepsilon_{nji} \langle \tau u_i {\bf  \nabla}_{m}
u_j \rangle ) / 2 ,$ and
$ \eta_{pm} = \eta \delta_{pm}  + \langle \tau u_p u_m  \rangle ,$
and $ \tau(r) $ is the scale-dependent momentum relaxation time, $ {\bf V}
= \langle {\bf u} \rangle ,$ the tensor $ I_{ij} $ is determined by an
external source of magnetic fluctuations, $ {\bf r} = {\bf y} - {\bf x} .$
We seek a solution for the second moment of the magnetic field in the
form
\begin{eqnarray}
\langle h_m({\bf x}) h_n({\bf x} + {\bf r}) \rangle &=&
W(r) \delta_{mn} + (r W' / 2) P_{mn}
\nonumber \\
&+& \mu(r) \varepsilon_{mnp} r_{p} / 2 \;,
\label{A22}
\end{eqnarray}
where $ P_{mn} = \delta_{mn} - r_m r_n / r^2 ,$ and $ W' = dW / dr .$
Note that the current helicity correlation function $ \mu(r) =
[\langle {\bf h}({\bf x}) \cdot (\vec{\bf \nabla} {\bf \times} {\bf h}
({\bf y})) \rangle + \langle {\bf h}({\bf y}) \cdot (\vec{\bf \nabla}
{\bf \times} {\bf h} ({\bf x})) \rangle] / 2 = \hat S \chi / 3 ,$ where
the magnetic helicity correlation function $ \chi(r)
= [\langle {\bf a}({\bf x}) \cdot {\bf h} ({\bf y}) \rangle
+ \langle {\bf h}({\bf x}) \cdot {\bf a} ({\bf y}) \rangle] / 2 ,$
and $ \hat S \chi = \chi'' + 4 \chi' / r .$
The correlation function $ \langle \tau u_m u_n \rangle $ is
\begin{eqnarray}
\langle \tau u_m({\bf x}) u_n({\bf x} + {\bf r}) \rangle &=& \eta_T
[ (F + F_c) \delta_{mn} + r F' P_{mn} / 2
\nonumber \\
&+& F'_c r_m r_n / r + \alpha(r) \varepsilon_{mnp} r_{p} / 2]
\label{B1}
\end{eqnarray}
(see \cite{EKRR97}), where $ \eta_T = u_0^2 \tau_0 / 3 $
is the turbulent magnetic diffusion, $ u_0 $ is the characteristic velocity
in the maximum scale $ l_0 $ of turbulent motions,
$ \tau_0 = l_0 / u_0 , \quad F(0)
= 1 - F_c(0) .$ The function $ F_c(r) $ describes the compressible
(potential) component, whereas $ F(r) $ corresponds to the vortical
part of the turbulence. The function $ \alpha(r) = -
[\langle \tau {\bf u}({\bf x}) \cdot (\vec{\bf \nabla} {\bf \times}
{\bf u} ({\bf y})) \rangle + \langle \tau {\bf u}({\bf y}) \cdot
(\vec{\bf \nabla} {\bf \times} {\bf u} ({\bf x})) \rangle]
(6 u_{0})^{-1} .$ In Eqs. (\ref{A22}) and (\ref{B1}) the
dimensionless distance $ r $ is measured in the units  $ l_0 .$

We use here the $ \delta $-correlated in time
random process to describe a turbulent velocity field.
Using the $ \delta $-correlated-in-time random process allows us to
obtain the analytical results for the anomalous scalings of the
two-point correlation functions of the magnetic field and magnetic
helicity. The results remain valid also for the velocity field
with a finite correlation time if the second-order correlation
functions of the magnetic field and magnetic helicity vary
slowly in comparison to the correlation time of the turbulent
velocity field (see, e.g., \cite{ZMR88}). We also take into
account the dependence of the
momentum relaxation time on the scale of the turbulent velocity field:
$ \tau ({\bf k}) = \tau_{0} k^{1-p} ,$ where $ p $ is the
exponent in spectrum of kinetic turbulent energy, $ k $ is the wave number
measured in the units $ l_{0}^{-1} .$

Using Eqs. (\ref{D30})-(\ref{B1}) we derive equations for the
correlation functions of the magnetic field $ W(t,r) $ and the
magnetic helicity $ \chi(t,r) .$ Indeed,
\begin{eqnarray}
\partial W / \partial t &=& (W'' + \zeta W' - \xi W) / m
\nonumber \\
&-& 2 (\alpha_{0} - \alpha(r)) \mu + \tilde I \;,
\label{R8} \\
\partial \mu / \partial t &=& \hat S [2 (\alpha_{0} - \alpha(r)) W +
\mu / m] \;,
\label{R8H}
\end{eqnarray}
where $ \alpha_{0} =
\alpha(r=0) = - \langle \tau {\bf u} \cdot (\vec{\bf \nabla} {\bf \times}
{\bf u}) \rangle / 3 $ is the $ \alpha $-effect, $ \tilde I $ is an
external source of magnetic fluctuations, and
$ 1 / m = 2 / {\rm Rm} + 2 [1 - F - (r F_c)'] / 3, $ and
$ \zeta = 4 / r + m (1 / m)' ,$ and $ \xi = 2 m (f' + 2 f'_{c}) / r ,$
and $ f = F + r F' / 3 ,$ and $ f_{c} = F_{c} + r F'_{c} / 3 $,
and $ {\rm Rm} = u_0 l_0 / \eta \gg 1 $ is the magnetic
Reynolds number, and the functions $ F(r) $ and $ F_c(r) $
are determined below.
Equation (\ref{R8}) and (\ref{R8H}) are written
in dimensionless variables: coordinates and time are measured
in the units  $ l_0 $ and $ \tau_0 ,$ the velocity is measured
in the units  $ u_0 ,$ the magnetic field is measured
in the units  $ B_0 .$ Note that in \cite{VK86} the system of
equations which is similar to Eqs. (\ref{R8}) and (\ref{R8H})
was derived. However, there are mistakes in the equations
derived in \cite{VK86}.
Since $ \mu(r) = \hat S \chi / 3 ,$ Eq. (\ref{R8H}) can be rewritten as
\begin{eqnarray}
\partial \chi / \partial t = (\chi'' + 4 \chi' / r) / m
+ 6 (\alpha_{0} - \alpha(r)) W \; .
\label{R9H}
\end{eqnarray}
Equation (\ref{R9H}) at $ r = 0 $ is given by $ (\partial \chi
/ \partial t)_{r=0} = 2(\chi'' + 4 \chi' / r)_{r=0} / {\rm Rm} .$
This implies that in a very high electrical conductivity limit
$ ({\rm Rm} \to \infty) $ the magnetic
helicity $ \chi(r=0) \equiv \langle {\bf a}({\bf x}) \cdot
{\bf h} ({\bf x}) \rangle $ is conserved.
We seek a solution of Eqs. (\ref{R8}) and (\ref{R9H}) in the form:
$ W (t,r) = (\Psi(r) \sqrt{m} / r^{2}) \exp (\Gamma t) $ and
$ \chi(t,r) = (\kappa (r) / r^{2}) \exp (\Gamma t) ,$
where the functions $ \Psi(r) $ and $ \kappa (r) $ are
determined by
\begin{eqnarray}
\Psi'' / m(r) - [\Gamma + U(r)] \Psi = v(r) [\kappa ''
- 2 \kappa  / r^{2}] / 9 m + I \;,
\label{R11} \\
\kappa '' / m(r) - [\Gamma + 2 / (m r^{2})] \kappa  = - v(r) \Psi \;,
\label{R11H}
\end{eqnarray}
and $ v(r) = 6 \sqrt{m} (\alpha_{0} - \alpha(r)) ,$ and
$ I = r^{2} \tilde I / \sqrt{m} ,$ and
$ U(r) = (\zeta^{2}  + 2 \zeta' + 4 \xi) / 4 m(r) .$
We consider the case of small magnetic Prandtl numbers $ {\rm Pr}_{m}
= \nu / \eta \ll  1 ,$ which is typical for many astrophysical
and geophysical applications, where $ \nu $ is the kinematic viscosity.
We choose the following model of turbulence.
Incompressible $ F(r) $ and compressible $ F_c(r) $ components
in the inertial range of turbulence $ r_d < r < 1 $ are given by
$ F(r) = (1 - r^{q-1}) / (1 + \sigma), \quad F_c(r) = (1 -
r^{q-1}) \sigma / (1 + \sigma) ,$
where $ \sigma $ is the degree of compressibility,
$ q $ is the exponent in
spectrum of the function $ \langle \tau u_m u_n \rangle ,$ and
$ r_d = {\rm Re}^{-1/(3-p)} , \quad p $ is the exponent in
the spectrum of kinetic turbulent energy, and $ {\rm Re} =
u_0 l_0 / \nu \gg 1 $ is the Reynolds number.
Note that the exponent $ p $ in the spectrum of kinetic turbulent
energy differs from that of the function  $ \langle \tau u_m
u_n \rangle $ due to the scale dependence of the momentum
relaxation time $ \tau $ of the turbulent velocity field.
The relation between $ p $ and $ q $ is $ q = 2p - 1 $
\cite{RK97}. Equation (\ref{R8}) for $ \alpha(r) = 0 $ and
$ \sigma = 0 $ was derived in \cite{K68}.

The solution of Eqs. (\ref{R11}) and (\ref{R11H}) can be obtained
using an asymptotic analysis (see, e.g.,
\cite{ZMR88,EKRR96,RK97,EKRR97}).
This analysis is based on the separation of scales. In particular,
the solutions of the Schr\"{o}dinger equations (\ref{R11})
and (\ref{R11H}) with a variable mass have
different regions where the form of the potential $ U(r) ,$ mass $
m(r) $ and, therefore, eigenfunctions $ \Psi(r) $ and $ \kappa(r) $
are different. Solutions in these different regions can be matched
at their boundaries. The results obtained by this asymptotic analysis
are presented below.

\section{Anomalous scalings}

We study a zero mode, i.e., we obtain the solutions of Eqs.
(\ref{R11}) and (\ref{R11H}) at $ \Gamma = 0 .$ In this case Eqs.
(\ref{R11}) and (\ref{R11H}) are given by
\begin{eqnarray}
\Psi'' / m(r) - \tilde U(r) \Psi &=& I \;,
\label{A1} \\
\kappa '' - 2 \kappa  / r^{2} &=& - v(r) m \Psi \;,
\label{A2}
\end{eqnarray}
where $ \tilde U(r) = U(r) - 4 m (\alpha_{0} - \alpha(r))^{2} ,$
and the function $ \alpha(r) = \alpha_{0} (1 - r^{q-1}) $ for $ 0 \leq
r \leq 1 ,$ and $ \alpha(r) = 0 $ for $ r > 1 ,$
and the external source of magnetic fluctuations $ \tilde I(r) = I_{0}
(1 - r^{s}) $ for $ 0 \leq r \leq 1 ,$ and $ \tilde I(r) = 0 $
for $ r > 1 ,$ and $ s > 0 .$ The solutions of Eqs. (\ref{A1})
and (\ref{A2}) have three characteristic regions.
In region I, i.e., for   $ 0 \leq r \leq  {\rm Rm}^{-1/(q-1)} ,$
the functions $ W(r) $ and $ \chi(r) $ are given by
$ W(r) = I_{\ast} (1 - \beta_0 {\rm Rm} \, r^{q-1}) ,$ and
$ \chi (r) = B_{1} + a_{1} \alpha_{0} {\rm Rm} \, r^{q+1} ,$
where $ \beta_0 = \beta_m + \xi_{0}
(q-1) / (q+2) ,$ and $ I_{\ast} \sim I_{0} {\rm Rm}^{(q+2)/2(q-1)} ,$
and $ a_{1} = A_{1} / [(q+1)(q+4)] ,$ and
$ \beta_m =  (1 + q \sigma) / 3 (1 + \sigma) ,$
and $ \xi_{0} = (1 + 2 \sigma) (2 + q) (q - 1) / 3 (1 + \sigma) .$
In region II, i.e.,
for  $ {\rm Rm}^{-1/(q-1)} \ll r \ll 1 ,$
the functions $ W(r) $ and $ \chi(r) $ are given by
\begin{eqnarray}
W(r) & = & A_2 m^{1/2} r^{-3/2} \cos(b \ln r + \varphi_{0} ) + W_{N} \;,
\label{L25} \\
\chi (r) & = & B_{2} + B_{3} / r^{3} + a_{2} \alpha_{0}
Re\{r^{-(q/2-1+ib)}\} \;,
\label{L1}
\end{eqnarray}
where $ W_{N} = - I_{0}  r^{3-q} / [2 \beta_{m} ((4 - q / 2)^{2}
+ \vert b \vert^{2})] ,$ and
\begin{eqnarray*}
b^{2} &=& \biggl( {q^2 - 4 \over 4 } \biggr) \biggl({ 3 + \sigma (4-q)
\over 1 + q \sigma} \biggr) \;,
\\
a_{2} &=& {(3 A_{2} / \sqrt{2}) \beta_{m}^{-3/2} \over [ ( (1 - q / 2)^{2}
+ \vert b \vert^{2}) ((4 - q / 2)^{2}
+ \vert b \vert^{2})]^{1/2} } \; .
\end{eqnarray*}
Note that when $ q \geq 2 $ the parameter $ b $ is a complex number
and $ Re\{r^{-(q/2-1+ib)}\} = r^{-q/2+1} \cos(b \ln r) ,$ and
when $ q < 2 $ the parameter $ b $ is a real number. For $ q < 2 $
the solution for $ W(r) $ is given by $ W(r) = m^{1/2} r^{-3/2} (A_{2}
r^{-\vert b \vert} + A_{4} r^{\vert b \vert} ) .$
In region III  $ (r \gg 1) ,$ the functions
$ W(r) = A_3 r^{-2} ( 3 \alpha_{0} \cos(3 \alpha_{0} r) -
r^{-1} \sin(3 \alpha_{0} r)) ,$ and
$ \chi(r) = B_{4} / r^{3} -6 A_{3} \alpha_{0} r^{-1}
\sin(3 \alpha_{0} r) ,$
where we take into account the boundary condition for the function
$ \chi(r) ,$ i.e., $ \chi(r) \to 0 $ for $ r \to \infty ,$
and a condition $ \chi(r) \to 0 $ for $ \alpha_{0} \to 0 .$
Matching the functions $ W(r) $ and $ W'(r) $ at the boundaries
of these regions yields $ A_{1} \sim A_{2} \sim A_{3} \sim A_{4} \sim
I_{0} .$ On the other hand,
matching the functions $ \chi(r) $ and $ \chi'(r) $ at the boundaries
of the regions yields $ B_{1} \sim I_{0} \alpha_{0}
{\rm Rm}^{(q-2)/2(q-1)} $ for $ 2 \leq q \leq 3 ,$ and $ B_{1} \sim
I_{0} \alpha_{0} {\rm Rm}^{(q-2+2b)/2(q-1)} $ for $ 1 < q  < 2 ,$
and $ B_{2} \sim I_{0} \alpha_{0} $
for $ 1 < q  \leq 3 ,$ and $ B_{3} \sim I_{0} \alpha_{0}
{\rm Rm}^{-(8-q)/2(q-1)} $ for $ 2 \leq q \leq 3 ,$ and $ B_{3} \sim
I_{0} \alpha_{0} {\rm Rm}^{(q-6+2b)/2(q-1)} $ for $ 1 < q  < 2 .$

The magnetic fluctuations are excited when the magnetic Reynolds number
$ {\rm Rm} > {\rm Rm}^{\rm cr} ,$ where the critical magnetic Reynolds
number $ {\rm Rm}^{\rm cr} $ is found in \cite{RK97}.
For incompressible fluid $ \sigma = 0 $
and $ p = 5/3 $ (Kolmogorov turbulence) the critical magnetic Reynolds
number $ {\rm Rm}^{({\rm cr})} = 412 ,$ while for compressible
fluid flow $ \sigma = 0.1 $ the value $ {\rm Rm}^{({\rm cr})} = 740 .$
For a larger parameter of compressibility
the critical magnetic Reynolds number increases sharply up to
$ {\rm Rm}^{({\rm cr})} \sim 10^{6} $ (see, \cite{RK97}).
First, we consider the case $ {\rm Rm} < {\rm Rm}^{\rm cr} ,$
i.e., when there is no self-excitation of the magnetic fluctuations.
In this case the magnetic fluctuations are sustained by an external source.
The first term in Eq. (\ref{L25}) for the correlation function of
the magnetic field $ W(r) $ in the inertial range
is given by $ W_{A} \sim r^{-q/2-1} \cos(b \ln r + \varphi_{0}) .$
This corresponds to the anomalous scaling
of the magnetic fluctuations. The normal scaling for the
second moment of the magnetic fluctuations is given by
the second term in Eq. (\ref{L25}): $ W_{N} \sim r^{3-q} .$
The general solution of equation for the second-order correlation function
of magnetic field $ W(r) $ includes solutions describing the anomalous
and normal scalings. The anomalous scaling
$ W_{A} \sim r^{-q/2-1} \cos(b \ln r + \varphi_{0}) $
can be presented as the real part of the power-law function
$ r^{\epsilon} $ with the complex exponent $ \epsilon = - q/2 - 1
+ i b(\sigma,q) .$
This anomalous scaling corresponds to the deviation from
the condition of the constant flux of magnetic fluctuations
over the spectrum. It describes the case $ 2 < q < 3 .$
When $ 1 < q < 2 $ the anomalous exponent in a low-Mach-number
compressible turbulent flow
is real, i.e., $ \epsilon = - q/2 - 1 + \vert b(\sigma,q) \vert .$
In the case of incompressible turbulent flow $ (\sigma = 0) $
and $ 1 < q < 2 $ this result coincides with that obtained in \cite{V96}.
For incompressible turbulent flow and $ 2 < q < 3 $
the anomalous scaling of magnetic field is the complex number
$ \epsilon = - q/2 - 1 - i b(\sigma=0,q) ,$ see \cite{RK97}.

Now we discuss solutions for the correlation function of the
magnetic helicity. Last term in Eq. (\ref{L1})
$ \propto Re\{r^{-(q/2-1+ib)}\} $ corresponds to the anomalous scaling
for the magnetic helicity correlation function.
The magnetic helicity $ \chi_{0} = \chi(r=0) $ in the limit of
very large electrical conductivity is conserved. This means
that the two-point correlation function of conserved property has
anomalous scaling. The reason for the anomalous scalings of the magnetic
helicity correlation function is that the magnetic field in the
equation for the two-point correlation function of
the magnetic helicity plays a role of a pumping with anomalous scalings
[see Eqs. (\ref{A1}) and (\ref{A2})].
This demonstrates a difference between
passive vector (magnetic field) and passive scalar. Indeed,
the quantity $ n^{2} $ (or $ T^{2}) $ in incompressible and
homogeneous turbulent fluid flow is conserved
(for infinitely small diffusivity or thermal conductivity). Corresponding
two-point correlation function has normal scaling.
On the other hand, the magnetic helicity is conserved and the
two-point correlation function of magnetic helicity has anomalous
scaling for a turbulent helical velocity field.
Thus we demonstrated here that two-point correlation function
of the conserved property does not necessarily have normal scaling.
Note that the nonlinear effects (i.e., self-consistent dynamics in
which the back-reaction of the Lorentz force is considered) are important
when the amplitude of the magnetic field is enough large,
i.e., when $ \langle {\bf h}^{2} \rangle
/ 4 \pi \sim \langle \rho {\bf u}^{2} \rangle .$ In this section
we consider the case when there is no magnetic dynamo, i.e., the magnetic
fluctuations are sustained by an external source $ I .$ The obtained
results are valid when the external source $ I $ is not very strong,
i.e., $ I \tau / 4 \pi \ll \langle \rho {\bf u}^{2} \rangle ,$
and the nonlinear effects are not important.

\section{Dynamics of magnetic helicity}

Now, we consider the case $ {\rm Rm} > {\rm Rm}^{\rm cr} ,$ i.e.,
when the magnetic fluctuations with zero mean magnetic field are
excited. What is the dynamics of the magnetic helicity in a
helical turbulent fluid flow? Equations (\ref{R11}) and
(\ref{R11H}) can be rewritten in the form $ (\hat Q + \hat V) {\bf
X} = \Gamma {\bf X} ,$ where $ {\bf X} $ is the vector-column with
the components $ X_{1} = \Psi ,$ and $ X_{2} = \kappa ,$ and the
matrix $ \hat Q \equiv Q_{ij} = 0 ,$ when $ i \not= j ,$ and $
Q_{11} = (d^{2} / dr^{2} - U) / m ,$ and $ Q_{22} = (d^{2} /
dr^{2} - 2 / r^{2}) / m ,$ and the matrix $ \hat V \equiv V_{ij} =
0 ,$ when $ i = j ,$ and $ V_{12} = - (d^{2} / dr^{2} - 2 / r^{2})
/ 9m ,$ and $ V_{21} = 1 .$ We consider the modes which satisfy
the following property: the change $ \alpha(r) \to - \alpha(r) $
in Eqs. (\ref{R11}) and (\ref{R11H}) results in the change $
\kappa(r) \to - \kappa(r) $ and $ \Psi(r) \to \Psi(r) .$ We seek a
solution of this equation in the form $ {\bf X} =
\sum_{k=1}^{\infty} x_{k} {\bf e}_{k} + \int y(\gamma) {\bf
E}(\gamma) \,d \gamma ,$ where the eigenfunctions $ {\bf e}_{k} $
and $ {\bf E}(\gamma) $ satisfy to the equations: $ \hat Q {\bf
e}_{k} = \gamma_{k} {\bf e}_{k} ,$ and $ \hat Q {\bf E}(\gamma) =
\gamma {\bf E}(\gamma) .$ Here $ {\bf e}_{k} $ is the
vector-column with the components $ e_{1k} = \tilde \Psi_{k} ,$
and $ e_{2k} = 0 ,$ and $ {\bf E}(\gamma) $ is the vector-column
with the components $ E_{1} = 0,$ and $ E_{2} = \tilde
\kappa(\gamma) .$ The functions $ \tilde \Psi_{k}(r) $ and $
\tilde \kappa(r) $ are determined by Eqs. (\ref{R11}) and
(\ref{R11H}) with the condition $ \alpha (r) = 0 .$ The equation
for the function $ {\bf e}_{k} $ describes nonhelical component of
the magnetic field correlation function and it has discrete
spectrum. On the other hand, the equation for the function $ {\bf
E}(\gamma) $ determines helical component of the magnetic field
correlation function and it has continuous spectrum. The
continuous spectrum corresponds to $ \gamma < 0 ,$ i.e., it
describes the relaxation of the magnetic helicity correlation
function. The discrete spectrum of the equation for the function $
{\bf e}_{k} $ corresponds to the generation of the magnetic
fluctuations $ (\gamma_{p} > 0) .$ The normalize conditions for
the eigenfunctions $ {\bf e}_{k} $ and $ {\bf E}(\gamma) $ are
given by $ \int m(r) {\bf e}_{k}^{\dag} {\bf e}_{p}\,dr = T(k)
\delta_{kp} ,$ and $ \int m(r) {\bf E}^{\dag}(\gamma) {\bf
E}(\gamma') \,dr = S(\gamma) \delta(\gamma - \gamma') . $ The
standard procedure used in quantum mechanics (see, e.g.,
\cite{LL75}) yields the equations for the functions $ x_{p} $ and
$ y(\gamma) ,$ i.e., $ (\gamma_{p} - \Gamma) x_{p} = L({\bf y}) ,$
and $ (\Gamma - \gamma) y(\gamma) = N({\bf x}) ,$ where $ N({\bf
x}) = (1 / S(\gamma)) \int v(r) m(r) \tilde \kappa(r) \tilde x(r)
\, dr ,$ and $ L({\bf y}) = (\gamma /  T(p)) \int v(r) \tilde
\Psi_{k}(r) m(r) \tilde y(r) \, dr ,$ and $ \tilde x(r) =
\sum_{k=1}^{\infty} x_{k} \tilde \Psi_{k}(r) ,$ and $ \tilde y(r)
= \int y(\gamma') \tilde \kappa(\gamma') \,d\gamma' .$ Now we use
a perturbation theory (see, e.g., \cite{LL75}), i.e., we seek the
solutions of the above equations in the form of series $ Z =
Z^{(0)} + \varepsilon Z^{(1)} + \varepsilon^{2} Z^{(2)} + \ldots
,$ where $ Z = x_{p}; y; \Gamma .$ The small parameter is $
\varepsilon \sim {\rm Rm}^{-(q+2)/2(q-1)} $ (see below). The
perturbation theory yields $ \Gamma^{(2k+1)} = 0 ,$ and $
\Gamma^{(0)} = \gamma_{p} ,$ and $ \Gamma^{(2)} = - L({\bf
y}^{(1)}) / x_{p}^{(0)} ,$ and $ y^{(2k)} = 0 ,$ and $ y^{(1)} =
N({\bf x}^{(0)}) / (\gamma_{p} - \gamma) ,$ and $ x_{p}^{(2k+1)} =
0 , $ and $ x_{p}^{(0)} = 1 ,$ and $ x_{p}^{(2)} $ is determined
by equation $ N({\bf x}^{(2)}) (\gamma_{p} - \gamma) =
\Gamma^{(2)} y^{(1)} ,$ etc.

For nonhelical turbulence $ [\alpha (r) = 0] $ the helical $ \mu(r) $ and
nonhelical $ W(r) $ parts of the magnetic field correlation function are
decoupled. This implies that the magnetic helicity can only relaxate
from the initial value.
In helical turbulence $ [\alpha (r) \not= 0] $ the magnetic helicity
$ \chi(r) $ depends on $ W(r) .$ This implies that the eigenfunction
$ {\bf e}_{p} $ of the discrete spectrum is modified, i.e. it has
spiral and nonspiral components: $ e_{1p} = \Psi_{p} $ and $ e_{2p} =
\kappa_{p} ,$
where $ \kappa_{p}(r) = \int_{0}^{\infty} \tilde \kappa(-\lambda,r)
V_{p}(\lambda) / [\gamma_{p} + \lambda] \,d\lambda +
O(\varepsilon^{3}) ,$
and $ \lambda = \vert \gamma \vert ,$ and $ \Psi_{p} = \tilde \Psi_{p}
+ O(\varepsilon^{2}) ,$ and $ V_{p}(\lambda) =
(1 / S(-\lambda)) \int v(r') m(r') \tilde \kappa(-\lambda,r)
\Psi_{p}(r') \, dr' .$
Using this equation we calculate the magnetic helicity $ \chi(r=0) .$
The result is given by
$ \chi(r=0) \sim - 4 \alpha_{0} \tau_{0} {\rm Rm}^{-(q+2)/2(q-1)} W(r=0) ,$
where $ W(r=0) = W_{0} \exp(\gamma_{p}t) .$ For nonhelical turbulence
there is only the relaxation of the initial magnetic helicity
$ \chi(t=0) \not= 0 .$
On the other hand, for helical turbulence the magnetic helicity
is excited due to the growth of magnetic fluctuations.
However, the magnitude of the magnetic helicity is very small
even if the hydrodynamic helicity is large, i.e.,
$ \chi(r=0) = - 4 \alpha_{0} \tau_{0} W(r=0) / {\rm Rm}^{13/10} ,$
where we use the Kolmogorov spectrum $ q = 5/3 .$
The realisability condition $ M(k) \geq \vert \chi(k) \vert k $
(see, e.g., \cite{M78}) allows us to estimate the maximum
possible value of the
magnetic helicity: $ \chi_{\rm max} \sim l_{0} W {\rm Rm}^{-1/(q-1)} ,$
where $ \langle {\bf h}^{2} \rangle = \int M(k) \,dk .$ Therefore,
$ \chi(r=0) / \chi_{\rm max} \sim (\alpha_{0} / u_{0})
{\rm Rm}^{-q/2(q-1)} \ll 1 .$ This means that the magnetic helicity with
zero mean magnetic field is very small even if the hydrodynamic helicity
is large. Only nonzero mean magnetic field can create large magnetic
helicity (see, e.g., \cite{ZRS83}).

\begin{acknowledgments}
We thank Dmitry Sokoloff for stimulating discussions
and valuable suggestions.
\end{acknowledgments}

\end{document}